\documentstyle[preprint,aps,psfig]{revtex} 

\begin{document}
\draft
\title{Statistical Analysis of Genealogical Trees for Polygamic Species.}
\author{Paolo De Los Rios$^1$ and Oscar Pla$^2$}
\address{$^1$Institut de Physique Th\'eorique, 
Universit\'e de Fribourg, CH-1700, Fribourg, Switzerland.}
\address{$^2$Instituto de Ciencia de Materiales, Consejo Superior de
Investigationes Cientificas, Cantoblanco, E-28049 Madrid, Spain}
\date{\today}
\maketitle

\begin{abstract}
Repetitions within a given genealogical tree provides some information
about the degree of consanguineity of a population. They
can be analyzed with techniques usually employed in
statistical physics when dealing with fixed point transformations.
In particular we show that the tree features 
strongly depend on the fractions of males and females in the population, 
and also on the offspring probability distribution. We check different
possibilities, some of them relevant to human groups, and compare them 
with simulations. 
\end{abstract}

One of the main problems encountered in the efforts to preserve
species from extinction is genetic diversity. Indeed, besides
environmental threats to the welfare of a species, a less obvious
but nonetheless extremely important issue is related
to the largeness of the genetic pool from which the genes of an
individual are taken. Such a problem is related to the degree of 
consanguinity within the population: the more relatives mate
among themselves, the more reduced is the genetic diversity of the population.
There are examples in the wilderness 
of species with a relatively small genetic variety:
from molecular biology it is known that cheetahs, for example,
show a high degree of consanguinity, probably due to some bottleneck
in the number of individuals in their population some ten thousands
of years ago; in human societies, it is well known that
high rank aristocrats in Europe kept marrying only among
themselves. As a consequence, the appearance of a hemophiliac
individual spread the genetic disease all over the reigning houses
of Europe. This example sheds light on the relevance of the genetic
diversity of a population for its conservation: species with a small
genetic pool are weaker against genetic diseases.
The above examples show that genetic redundancy can come as a consequence of
a reduced population. 

In this Letter we address the same problem
from a different (but we believe complementary) standpoint:
we are interested in the genealogical trees of individuals
of species where the male-to-female ratio is not $1$ as in humans
(here we define this ratio taking into account only males and females that
are sexually mature). Among
such examples we can name lions, sea lions and some antelopes, 
where each successfully reproducing male mates with more than one female
(similar arguments could also be applied to polygamic human groups). 
Extreme cases are insects like bees and termites,
where for every reproductive female (queen) 
there are very many males.

We measure the genetic redundancy in the gene pool of an individual by
measuring the number of times that one of its ancestors many generations
in the past appears more than once in its genealogical tree.
Indeed, if no relatives would mate among themselves then, since every individual
has a mother and a father, it would have $2^g$ ancestors $g$ generations
in the past, half of them males and half of them females. Each of them
would appear
only once in the genealogical tree of their present descendents. Going back
some tens of generations in the past,
the number of ancestors would largely exceed the population itself. 
The only way out from this paradox is to
assume that relatives indeed mate among themselves. As a consequence
some individuals appear more than once in the genealogical tree of their
descendents (that is, more than one branch of the tree had origin
from such individuals), thus reducing the genetic pool from which their
genes are taken.

We take a population of $N$ individuals, and we assume that it does not
change in time.
There is a fraction $f N$ of males and $(1-f)N$ of females, ad also
this fraction remains
constant in time.
Every male mates therefore, on the average, with $1/f-1$ females.
Here in general we make the (politically uncorrect) assumption
that the fraction of males is less than $1/2$. Since in this model 
there is no difference between males and females, the opposite situation
is obtained with a transformation $f \to 1-f$ (everything is symmetric
with respect to $f=1/2$).
We apply and extend the same scheme as developed in \cite{DMZ99}, generalizing it
to the case of male fractions different from $1/2$.

Given an individual in the present generation, we are interested in the
number of times its ancestors at a previous generation $g$
appear in the genealogical tree of that individual
(at $g=1$ we find parents, at $g=2$ the grandparents, and so on).
We therefore define $m_r(g)$ ($f_r(g)$) as the number of males (females)
appearing $r$ times at generation $g$ 
in the genealogical tree of an individual at generation $0$, the
present one.

The normalization of $m_r(g)$ and $f_r(g)$ implies that we can write
\begin{equation}
\sum_{r=0}^{\infty} m_r(g) \Delta r = f N \;,\;\;\;
\sum_{r=0}^{\infty} f_r(g) \Delta r = (1-f) N  
\label{normalization of number}
\end{equation}
where $\Delta r = 1$ trivially (but it is useful to write it explicitly
for future rescalings). 
Since an individual at generation $0$ has $2^{g-1}$ male ancestors
(not necessarily distinct) 
at generations $g$ (and $2^{g-1}$ female ancestors as well),
we can also write
%
%
%
\begin{equation}
\sum_{r=0}^{\infty} r m_r(g) \Delta r = \sum_{r=0}^{\infty} 
r f_r(g) \Delta r = 2^{g-1}  \;\;\;.   
\label{summed repetitions}
\end{equation}
We define then the probabilities connected to $m_r(g)$ and $f_r(g)$.
These are probabilities defined over the population at generation $g$.
Therefore we have
\begin{equation}
M_r(g) = \frac{m_r(g)}{f N} \;,\;\;\;
F_r(g) = \frac{f_r(g)}{(1-f)N}  \;.     
\label{probabilities}
\end{equation}

Using (\ref{probabilities}) we rewrite (\ref{normalization of number}) as 
%
%
%
\begin{equation}
\sum_{r=0}^{\infty} M_r(g) \Delta r = \sum_{r=0}^{\infty} F_r(g) 
\Delta r = 1  
\label{intermediate1}
\end{equation}
and (\ref{summed repetitions}) as 
\begin{equation}
\sum_{r=0}^{\infty} r M_r(g) \Delta r = \frac{2^{g-1}}{f N} \;,\;\;\;
\sum_{r=0}^{\infty} r F_r(g) \Delta r = \frac{2^{g-1}}{(1-f)N}\;.   
\label{intermediate2}
\end{equation}
Finally we rescale $r$, $F_r(g)$ and $M_r(g)$ as
\begin{eqnarray}
P_M(r,g) = \frac{2^{g-1}}{f N} M_r(g) \;&,&\;\;\;
P_F(r,g) = \frac{2^{g-1}}{(1-f) N} F_r(g) \nonumber \\
w_M(g) = \frac{f N}{2^{g-1}} r \;&,&\;\;\;
w_F(g) = \frac{(1-f) N}{2^{g-1}} r \;\;\;.
\label{rescaled variables}
\end{eqnarray}
With these definitions Eqs.(\ref{intermediate1}) become
%
%
%
\begin{equation}
\int_0^\infty P_M(w_M,g) d w_M =
\int_0^\infty P_F(w_F,g) d w_F = 1    
\label{new rescaled probability}
\end{equation}

and Eqs.(\ref{intermediate2}) become
%
%
%
\begin{equation}
\int_0^\infty w_M P_M(w_M,g) d w_M = 
\int_0^\infty w_F P_F(w_F,g) d w_F = 1  \;\;\;.   
\label{new rescaled first moments}
\end{equation}

From (\ref{new rescaled probability}) we see that $P_M(w_M,g)$ and 
$P_F(w_F,g)$ can be considered true probabilities.
Next, we can write a system of equations for $w_m(g)$ and $w_F(g)$.
A male $i$ at generation $g+1$ in the past has a number of repetitions 
that is given
by the number of repetitions of his children at generation $g$.
Therefore
\begin{equation}
r_{M,i}(g+1) = \sum_{j\,son\,of\,i} r_{M,j}(g) +  \sum_{j\,daughter\,of\,i} r_{F,j}(g)
\label{male equation}
\end{equation}
and analogously for females
\begin{equation}
r_{F,i}(g+1) = \sum_{j\,son\,of\,i} r_{M,j}(g) +  \sum_{j\,daughter\,of\,i} r_{F,j}(g)
\label{female equation}
\end{equation}
%

%
%
Dividing the first equation for $2^{g-1}/fN$ we get 
\begin{equation}
w_{M,i}(g+1) = \frac{1}{2}\sum_{j \,son \,of \,i} w_{M,j}(g) +  
\frac{f}{2(1-f)}\sum_{j \,daughter \,of\, i} w_{F,j}(g)\;\;.
\label{recurrence for w}
\end{equation}
Dividing (\ref{female equation}) for $2^{g-1}/(1-f)N$ we get the analogous
equation for females.

We assume a stable (on the average) population of $N$ 
individuals divided in two parts whose proportions are also (on the average)
stable. Therefore the number of sons (daughters) that an individual 
can have has to obey
well defined probability distributions. In our simulations we
proceed backward in time, keeping the population fixed at $N$ and the 
male proportion fixed at $f$. Since we assign to every individual
a couple of parents at random in the previous generation, the corresponding 
son/daughter probability distributions are binomials distributions.
More precisely, the probability that a male has $k$ sons is
\begin{equation}
p_{mm}(k) = \left( \begin{array}{c} fN \\ k \end{array} \right) \left(\frac{1}{fN}\right)^k
\left(1- \frac{1}{fN} \right)^{fN-k}
\label{pmm}
\end{equation} 
and that he has $k$ daughters is
\begin{equation}
p_{mf}(k) = \left( \begin{array}{c} (1-f)N \\ k \end{array} \right) 
\left(\frac{1}{fN}\right)^k
\left(1- \frac{1}{fN} \right)^{(1-f)N-k}
\label{pmf}
\end{equation} 
Analogous distributions can be written for 
$p_{ff}(k)$ and $p_{fm}(k)$.
%
%
%
%

We assume that the population is very large ($N\to \infty$)
and that all the $w$'s are independent (this is verified in the limit
of large $N$). 
In this limit the offpring probabilities 
become
%
%
%
\begin{eqnarray}
p_{mm}(k,f) &=& p_{ff}(k,f) =\frac{e^{-1}}{k!} \nonumber \\
p_{mf}(k,f) &=& p_{fm}(k,1-f) =\frac{e^{-(1-f)/f}}{k!} 
\left(\frac{1-f}{f} \right)^k\;\;\;.
\label{poissons}
\end{eqnarray}
In the case $f=1/2$ we recover the distributions used in \cite{DMZ99}.

Upon defining the generating functions
\begin{eqnarray}
G_g(\lambda) &=& \int_0^\infty e^{-\lambda w_M} P_M(w_M,g) d w_M \nonumber \\
H_g(\mu) &=& \int_0^\infty e^{-\mu w_F} P_F(w_F,g) d w_F  
\label{generating functions}
\end{eqnarray}
we find then that (\ref{recurrence for w}) become
\begin{eqnarray}
G_{g+1}(\lambda) &=& \sum_{k=0}^\infty \sum_{j=0}^\infty p_{mm}(k) \left[ G_g\left(\frac{\lambda}{2}\right)
\right]^k p_{mf}(j) \left[H_g\left(\frac{\lambda}{2} \frac{f}{1-f}\right) \right]^j \nonumber \\
H_{g+1}(\mu) &=& \sum_{k=0}^\infty \sum_{j=0}^\infty p_{fm}(k) \left[ G_g\left(\frac{\mu}{2} \frac{1-f}{f}\right)
\right]^k p_{ff}(j) \left[H_g\left(\frac{\mu}{2}\right)\right]^j 
\label{equations for G and H}
\end{eqnarray}
where also the equation for females has been written explicitely.

Substituing (\ref{poissons}) in (\ref{equations for G and H}) we get, 
after some algebra,
\begin{eqnarray}
G_{g+1}(\lambda) &=& \exp\left[-\frac{1}{f} + G_g\left(\frac{\lambda}{2}\right) + \frac{1-f}{f} 
H_g\left( \frac{\lambda}{2}\frac{f}{1-f} \right) \right] \nonumber \\
H_{g+1}(\mu) &=& \exp\left[-\frac{1}{1-f} + \frac{f}{1-f} G_g\left(\frac{\mu}{2} \frac{1-f}{f}\right) +  
H_g\left( \frac{\mu}{2}\right) \right] 
\label{resummed equations}
\end{eqnarray}

These equations are clearly symmetric in $f \to 1-f$, since we 
do not make any distinction
between males and females apart from the male proportion $f$.

Next, we analyze the stationary equations, $g=\infty$,
\begin{eqnarray}
G(\lambda) &=& \exp\left[-\frac{1}{f} + G\left(\frac{\lambda}{2}\right) + \frac{1-f}{f} 
H\left( \frac{\lambda}{2}\frac{f}{1-f} \right) \right] \nonumber \\
H(\mu) &=& \exp\left[-\frac{1}{1-f} + \frac{f}{1-f} G\left(\frac{\mu}{2} \frac{1-f}{f}\right) +  
H\left( \frac{\mu}{2}\right) \right] 
\label{stationary equations}
\end{eqnarray}

\begin{figure}
\centerline{\psfig{file=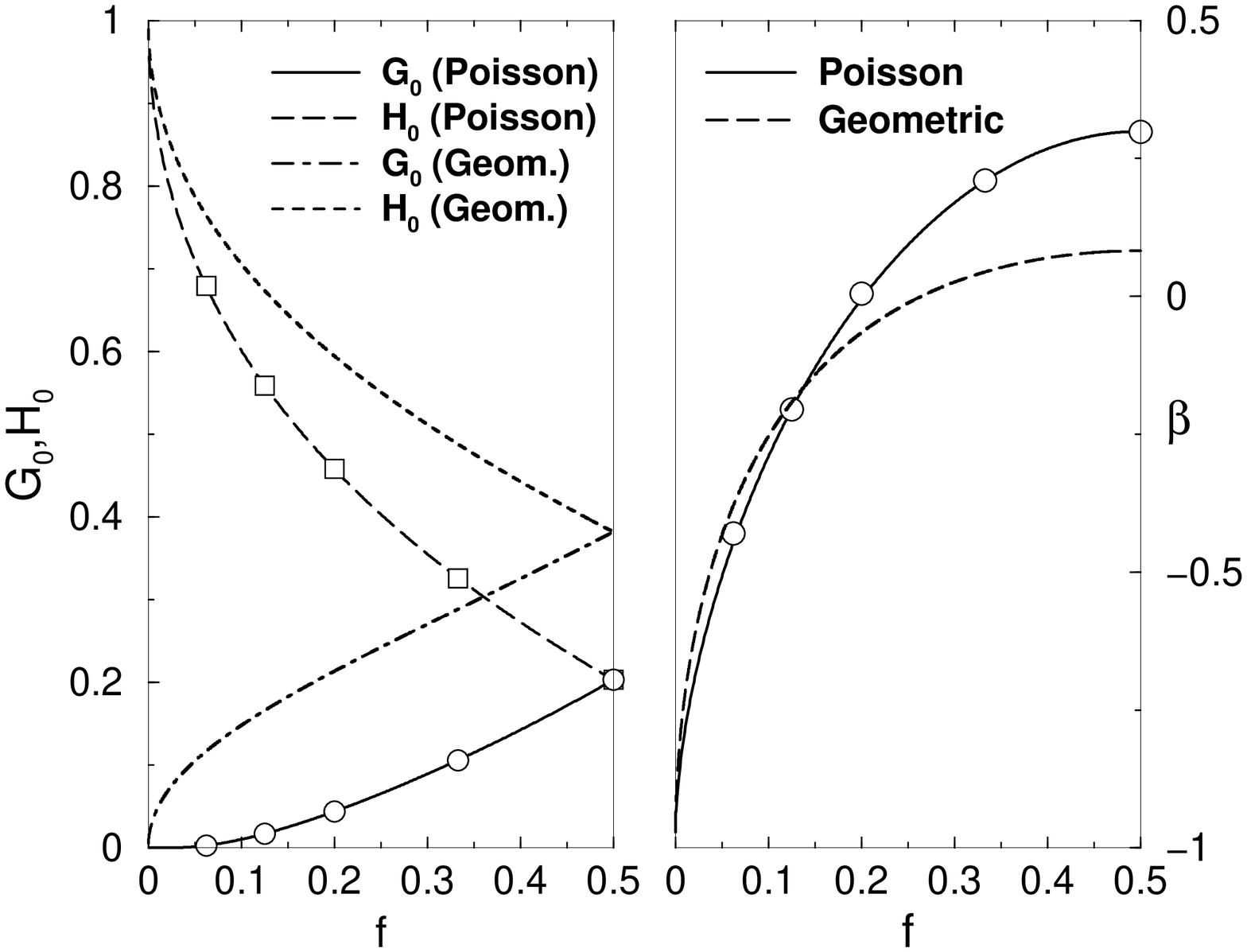,width=8.5cm,angle=0}} 
\caption{Left: Asymptotic fraction of males and females who 
do not belong to the genealogical tree
of a given individual in the present generation. Circles and squares are data
from simulations for $30$ generations over a population of $20000$
individuals, with (from right to left) $f=1/2,\;1/3,\;1/5,\;1/8,\;1/16$. 
Right: Exponent $\beta$ as a function of the fraction $f$ of males.} 
\label{Fig: fig1}
\end{figure}

The probability that a male (a female) in the past does not 
appear in the genealogical tree of
a given individual in the present generation is 
recovered sending $\lambda,\mu \to \infty$ (by tauberian theorems, the limit
$\lambda,\mu \to \infty$ corresponds to the limit $r_M, r_F =0$).
Therefore, upon calling $G_0 = G(\infty)$ and $H_0 = H(\infty)$ we have
\begin{eqnarray}
G_0 = \exp\left( -\frac{1}{f} + G_0 + \frac{1-f}{f} H_0 \right) \nonumber \\
H_0 = \exp\left( -\frac{1}{1-f} + \frac{f}{1-f} G_0 + H_0 \right)
\label{fixed point equations}
\end{eqnarray}
These equations can be solved numerically and the solution is shown in 
Fig.\ref{Fig: fig1}(Left) (the results of the simulations agree with
this solution up to the third significative digit). 

Next, we expand (\ref{stationary equations}) 
around the fixed point assuming that
$P_M(w_M) \sim G_0 \delta(w_M) + w_M^{\beta_M}$ and 
$P_F(w_F) \sim H_0 \delta(w_F)+ w_F^{\beta_F}$ for 
$w_M,w_F \to 0$, which translates, by
tauberian theorems, to
\begin{equation}
G(\lambda) = G_0 + A_M \lambda^{-\beta_M -1} \;,\;\;\;
H(\mu) = H_0 + A_F \mu^{-\beta_F -1} 
\label{asymptotics}
\end{equation}
for $\lambda,\mu \to \infty$.
Eqs.(\ref{stationary equations}) then become
\begin{eqnarray}
G_0 \left[ 2^{\beta_M+1}+ \frac{A_F}{A_M} \left(2 \frac{1-f}{f}\right)^{\beta_F+1} 
\lambda^{\beta_M-\beta_F} \right] &=& 1 \nonumber \\
H_0 \left[ 2^{\beta_F+1}+ \frac{A_M}{A_F} \left(2 \frac{f}{1-f}\right)^{\beta_M+1} 
\mu^{\beta_F-\beta_M} \right] &=& 1 
\label{expanded equations}
\end{eqnarray}

Eqs. (\ref{expanded equations}) are well defined only if $\beta_M=\beta_F=\beta$, 
and therefore we get, after some algebra,
\begin{equation}
2^{\beta+1} (H_0+G_0) = 1
\end{equation}
from which we can calculate the exponent $\beta$ as a function of $f$, shown in Fig.\ref{Fig: fig1}(Right).
%
%
 
From (\ref{fixed point equations}) it is also possible to get
the analytic behavior of $H_0$, $G_0$ and $\beta$ close to $f=0$:
\begin{equation}
G_0 \sim e^{-\sqrt{\frac{2}{f}}} \;,\;\;
H_0 \sim 1-\sqrt{2 f} \;,\;\; \beta \sim -1 + \frac{\sqrt{2}}{\ln 2} f^\frac{1}{2}
\label{f expansion}
\end{equation}
%
%
%

\begin{figure}
\centerline{\psfig{file=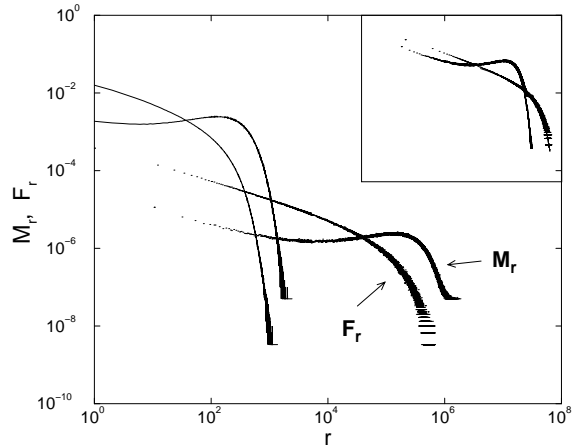,width=7.5cm,angle=0}} 
\caption{Male and female repetition probabilities after $20$ and $30$
generations (the latters are marked by arrows) for a male fraction $f=1/16$. 
In the inset we show the collapse of the rescaled distributions.} 
\label{Fig: fig3}
\end{figure}

As an exemple of distributions,
in Fig.\ref{Fig: fig3} we show $M_r(g)$ and $F_r(g)$, (\ref{probabilities}), 
and in the inset their rescaled counterpart according to 
(\ref{rescaled variables}), for $f=1/16$. The exponent $\beta$
is negative, as from our analytical calculations. The delta function
for $r=0$ has been omitted for scale reasons.

The dependence of $\beta$ from $f$ shows that such an exponent is highly
nonuniversal and that it is extremely sensitive to
the explicit form of the distributions (\ref{poissons}).
This becomes important when looking at real data. 
In the thirties Lotka\cite{Lotka} fitted the probability of a man to
have $k$ sons in the United States by a geometric distribution $p_k=b_{mm}c_{mm}^{k-1}$ for
$k\neq 0$, and $p_0=d_{mm}$, with $c_{mm}=0.5893$, $d_{mm}=0.4825$ and 
$b_{mm}$ chosen for normalization. 
Clearly, such a distribution is not a Poisson distribution as
used above. Moreover it would give a rate of increase in the population
of $N_{g}/N_{g+1}=1.26$.

Since in the definition of $P$ and $w$ in (\ref{rescaled variables}) depend on $g$, the particular
value of $N_g$ can be explicitly incorporated in it. The left
hand side of (\ref{recurrence for w}) is now multiplied by
$N_g/N_{g+1}$. The probabilities for a male to be son of a male and a
female to be daughter of a female will be those of Lotka, and the other ones
can be evaluated by maintaining the fraction of males and females in the
population constant, which translates in the constraints:
\begin{equation}
\frac{1-d_{mf}}{1-c_{mf}} = \frac{1-f}{f} \frac{N_g}{N_{g+1}} \;,\;\;\;
\frac{1-d_{fm}}{1-c_{fm}} = \frac{f}{1-f} \frac{N_g}{N_{g+1}}\;. 
\label{mixmax}
\end{equation}
We can then rewrite (\ref{resummed equations}) as
\begin{eqnarray}
\frac{N_g}{N_{g+1}}G_{g+1}(\lambda) &=& \left(d_{mm}+\frac{(1-c_{mm})(1-d_{mm})
G_g(\lambda/2)}
{1-c_{mm}G_g(\lambda/2)}\right)\left(d_{mf} +\frac{(1-c_{mf})(1-d_{mf})
H_g(\lambda/2)}
{1-c_{mf}H_g(\lambda/2)}\right)
\nonumber \\
\frac{N_g}{N_{g+1}}H_{g+1}(\mu) &=&
\left(d_{fm}+\frac{(1-c_{fm})(1-d_{fm})G_g(\mu/2)}{1-c_{fm}G_g(\mu/2)}\right)
\left(d_{ff}+\frac{(1-c_{ff})(1-d_{ff})H_g(\mu/2)}{1-c_{ff}H_g(\mu/2)}\right)
\end{eqnarray}
Here we examine two different cases. First, we take $f=1/2$ and all $c$s and
$d$s as from Lotka. We find that the probability $G_0=H_0=0.231$, different
from the one obtained with Poisson distributions\cite{DMZ99}.
Then we impose that the population size remains constant, $N_g=N_{g+1}$,
but allow for different male fractions. Moreover, for simplicity,
we choose $d=1-c$ for the four probability distributions, 
in such a way that they
become genuine geometric distributions:
$p_{mm}(k)=p_{ff}(k) = 1/2^{k+1}$, $p_{mf}(k) = f (1-f)^k$ and
$p_{fm}(k) = (1-f) f^k$. The results for $G_0$ and $H_0$ are shown also in 
Fig.\ref{Fig: fig1}(Left). The exponent $\beta$ is shown in Fig.\ref{Fig: fig1}(Right).
$G_0$ and $H_0$ approach their limit for $f\to 0$ as $f^{1/2}$.
In particular, the values for $f=1/2$ are clearly different from the ones
with Poisson distributions\cite{DMZ99}. 
We find therefore that neither $G_0$ and $H_0$, nor $\beta$ are universal,
although their behavior with respect to $f$ does not, qualitatively,
depend on the details of the chosen offspring distribution.
Actually, the relevance of the distribution to be used is
hardly overestimated: one should take distributions obtained from the analysis
of real data, in order to draw more detailed conclusions\cite{nota poisson}.

The present results show that, besides bottlenecks in the population size,
there may be other factors affecting the largeness of the genetic pool from
which the genes of an individual are taken. 
Indeed, for species with a very low value of $f$ we find that most females
do not contribute to the genes of an individual in the present
generation, whereas most males (who are anyway a little fraction $f$
of the entire population) do. 
As an extreme case (and exchanging males with females), 
in the absence of interbreeding between different hives, a single 
bee queen gives its genes to all subsequent generations. Some genetic
mutation will become rapidly a genetic trait of the whole progeny.
In case of bad mutations, they could well wipe out the whole family
line. 
Although not dangerous {\it per se}, since bees and alike
are extremely numerous, such a feature can make
the species more sensitive to population size fluctuations.

In conclusion, we have generalized and analyzed the model proposed in
\cite{DMZ99} to the realistic case of species and human groups with
male-to-female mating ratios different from $1$. Our results point
out that the genes of an individual are taken from a pool
whose largeness strongly depends on the male-to-female ratio, with 
important consequences when the population size strongly fluctuates.
We are currently investigating the coupling effects between these
different factors.
Yet our results, although qualitatively of a general applicability,
clearly show that quantitative estimates can only come when 
the analytical treatment is implemented with field data, since, as it
is evident from Figs.\ref{Fig: fig1}(Left) and \ref{Fig: fig1}(Right), different offspring
probability distributions give rise to different quantitative results.
This is a highly non-universal problem.

We thank F. Guinea for useful comments and discussions.
P. De Los Rios thanks the Instituto de Ciencia de Materiales in Madrid,
where this work was begun, for its kind hospitality.
This work has been partially supported by the European Network 
contract FMRXCT980183.

\begin{thebibliography}{99}
\bibitem{nota poisson} 
Indeed, Poisson distributions such as (\ref{poissons}) are the natural
candidate matching efficient simulations in the large $N$ limit, but once
this limit has been taken, we are free to use whatever distribution we like.
\end{thebibliography}
\end{document}